\input harvmac
\input tables

\Title{hep-th/yymmnn RU-96-27, SCIPP 96/15}
{\vbox{\centerline{Couplings and Scales in Strongly
Coupled}
\centerline{ Heterotic String Theory}}}
\bigskip
\centerline{Tom Banks}
\smallskip
\centerline{\it Department of Physics and Astronomy}
\centerline{\it Rutgers University, Piscataway, NJ 08855-0849}
\smallskip
\centerline{Michael Dine}
\smallskip
\centerline{\it Santa Cruz Institute for Particle Physics}
\centerline{\it University of California, Santa Cruz, CA   95064}
\bigskip
\baselineskip 18pt
\noindent
If nature is described by string theory, and if the
compactification radius is large (as suggested by
the unification of couplings), then the theory is in a regime
best described by the low energy limit of $M$-theory.
We discuss some phenomenological aspects
of this view. The scale at which conventional quantum field theory
breaks down is of order the unification scale and consequently 
(approximate)
discrete symmetries are essential to prevent proton decay. There
are one or more light axions, one of which solves the strong CP problem.
Modular cosmology is still problematic but much more complex than
in perturbative string vacua.
We also consider a range of more theoretical issues, focusing
particularly on the question of stabilizing the moduli.  We
give a simple, weak coupling derivation of Witten's expression
for the dependence of the coupling constants on the eleven
dimensional radius.
We discuss the criteria for the validity of the
long wavelength analysis and find that the ``real world" seems
to sit just where this analysis is breaking down.  On the other
hand, residual constraints from $N=2$ supersymmetry
make it difficult to see how the moduli can be stabilized
while at the same time yielding a large hierarchy.

\Date{4/96}

\def\Mfour{M_4}
\def\Meleven{M_{11}}
\def\lp{l_{11}}
\def\Releven{R_{11}}
\newsec{Introduction and Summary}

The only vacuum independent 
quantitative predictions of weakly coupled heterotic string
phenomenology are a relation between the Planck mass,
the four dimensional gauge coupling and the string tension,
and a relation between the unification scale and the
scale of compactification.  This last prediction is rather
troubling.  For if we suppose that the successful supersymmetric
unification of couplings is not an accident, then one predicts
that the scale of compactification is a factor of $20$ or so below
the string scale.  This, in turn, implies that the dimensionless
coupling of string theory is of order $10^7$, so that a weak
coupling description surely does not make sense\foot{These statements
are valid for more or less isotropic Calabi-Yau manifolds.  In
\ref\banksdine{T. Banks and M. Dine,
``Coping with Strongly Coupled String Theory,''
hep-th/9406132, Phys. Rev. {\bf D50}
(1994) 7454. } we argued that highly anisotropic manifolds could
resolve this problem. There have also been attempts to extract
conventional four dimensional unified gauge theories from string theory.}.

On the other hand, it has been argued for many years that
string theory cannot be weakly coupled if it describes nature.
In the weak coupling region, 
the dilaton potential almost certainly cannot be
stabilized\ref\dineseiberg{
M. Dine and N. Seiberg,
Phys. Lett. {\bf 162B}, 299 (1985),
and in {\it
Unified String Theories}, M. Green and D. Gross, Eds. (World Scientific,
1986).}.
So perhaps we should simply accept the facts as they appear,
and suppose that the compactification scale, $R$, is large,
in heterotic string tension units,
and the theory is strongly coupled.  One might worry that,
by duality, such a strong coupling region would be mapped
into a weakly coupled region of some other string theory
(or of $M$ theory) and that this region would suffer from
some version of the dilaton runaway problem.  However,
in ref. \ref\truly{M. Dine and Y. Shirman,``Truly Strong Coupling
and Large Radius in String Theory,''
SCIPP-96-07, hep-th/9601175}, it
was pointed out that all of the known dualities map the
region of large radius, strong coupling, and fixed four dimensional
coupling to other strongly coupled theories (or at least
theories in which the couplings are not arbitrarily small).

Witten has recently taken this viewpoint to its logical
conclusion\ref\wittenstrong{E. Witten, ``Strong Coupling Expansion of
Calabi-Yau Compactification, hep-th/\break 9602070.}.  At strong
coupling, the heterotic theory is described, at low energies,
by $11$-dimensional supergravity. More generally,
the strong coupling limit of the theory has been
called $M$-theory\ref\mtheory{
M.J. Duff,
P. Howe, T. Inami and K.S. Stelle,
Phys. Lett. {\bf 191B} (1987) 70;
C.M.Hull, P.K.Townsend, ``Unity of
Superstring Dualities'', {\it Nucl. Phys.} {\bf B438}, (1995),409,
hep-th/9410167; E. Witten,
``String Theory Dynamics in Various Dimensions," Nucl.
Phys. {\bf B443} (1995) 85, hep-th/9503124; J. Schwarz,
``The Power of M Theory," Phys. Lett. {\bf B367}
(1996) 97, hep-th/9510086;
P. Horava and E. Witten, ``Heterotic and Type I String Dynamics From
Eleven Dimensions'', {\it Nucl. Phys.} {\bf B460}, (1996),506, hep-th/9510209;
``Eleven-Dimensional Supergravity on a Manifold with
Boundary," hep-th/9603142.}.
  Witten
has argued that  M-theory might well provide a
better description of nature than weakly coupled strings.
The $M$-theory description is valid, as we will
see shortly, when a certain parameter, which we will call
$\epsilon$,  is small.
This parameter seems to be of order
one in the real world, so the $M$ theory
description is likely to be at least qualitatively
much better than the weak coupling string description.

Compactification of the $E_8 \times E_8$ heterotic
theory on a Calabi-Yau space, ${\bf X}$,
is dual to $M$ theory compactified on ${\bf X} \times {\bf S^1
/ Z_2}$.
Using formulas presented in \wittenstrong,
one finds the following connections between the
$11$ dimensional Planck mass, $M_{11}$ (defined in terms of the
coefficient of the Einstein lagrangian in 11 dimensional supergravity, as
$M_{11}= \kappa_{11}^{-2/9}$), the $11$-dimensional radius,
$\Releven$, and the compactification radius, $R= V^{1/6}$, where
$V$ is the volume of the Calabi-Yau space on the boundary with unbroken
$E_6$ gauge group:
\eqn\rhosquared{\Releven^2 =  {\alpha_{GUT}^3 V \over 512 \pi^4 G_N^{2}
 },}
where $G_N$ is the four dimensional Newton's constant;
\eqn\meleven{M_{11}= R^{-1} \left (2 (4 \pi)^{- 2/3} \alpha_{GUT}
\right )^{-1/6}.}

Substituting reasonable
phenomenological values, one finds that the eleven dimensional
Planck length is roughly half the compactification radius,
while the eleven dimensional radius is about ten times the
compactification scale!   So one might hope that
eleven (or five)-dimensional supergravity provides
at least a crude approximation to the real
world.  Moreover, if this viewpoint
is correct, dramatic new physics occurs long
before one encounters the four dimensional Planck scale.  The
universe first looks five dimensional, then eleven dimensional.
The four dimensional Planck scale, $\Mfour$, is simply
a parameter of low energy physics; there is no
interesting new dynamics at this scale! {\it Quantum Gravitational (more
properly Quantum M theoretical) effects, become important at the
unification scale}.
This has possible implications for many
questions, including issues of early universe cosmology.
These are among the issues we will explore in this paper.

The qualitative physics of this {\it M theory regime}
 is quite different from that of
weakly coupled heterotic strings, which are no longer the lowest energy
excitations.  The fact that the compactification scale is large in
string tension units is a consequence of the fact that heterotic strings are
membranes stretched between the two walls of the eleven dimensional
world.  The fundamental energy scale in this regime is the eleven
dimensional Planck mass $\Meleven$.  The membrane tension is one in these
units but the heterotic membrane is large because the eleventh dimension
is an order of magnitude larger than $\lp=\Meleven^{-1}$.
The heterotic string tension, $T_h$ is $\Meleven^3 \Releven$.
The compactification radius is of order one in $\lp$ units, and this is
what determines the unification scale.

While the $M$-theory description should be qualitatively
much better than the weak coupling string description,
the universe is probably not
in a regime where one can simply compute in the
classical, low energy eleven dimensional supergravity
theory.  In the
classical supergravity theory, the expansion parameter
is 
\eqn\smallparameter{\epsilon =\kappa^{2/3}\Releven/R^4.}
This number is of
order one.  So we might expect unknown quantum
$M$-theory corrections to be of order one.  (This should be
compared with
the situation in the weakly
coupled string theory description, where
the ``small parameter" is of order $10^7$.)  As we will
discuss, this is just as well, since in the weak coupling
limit one could not understand the stabilization of the moduli.

Even before exploring any detailed dynamics, the view that the
universe is approximately five dimensional
has interesting consequences.  Consider, for example, the
strong CP problem.  It is well known that in four dimensional
string models, there is always a ``model-independent" axion,
the partner of the usual dilaton, with the potential to solve
the strong CP problem.  The superpartners of the Kahler, $(1,1)$,moduli of X
provide other axion candidates.
In weakly coupled string theory,
 world-sheet instanton effects break the Peccei-Quinn (PQ)
symmetries of these Kahler axions, 
by amounts of order $e^{-{R^2}}$.  Usually, since
$R$ is assumed to be a number of order one in $T_h$ units, these breaking
effects are also taken to be of order one.  However, if $R$ is
large, this factor can be extremely small. Indeed, in the five
dimensional picture, these axions lie in vector multiplets and the
associated PQ symmetries are \lq\lq would-be five dimensional gauge
symmetries" that are broken only by boundary effects and by membrane
instantons.  The latter are highly suppressed because of the large size
of the membranes (actually, because it occurs in a holomorphic
superpotential this effect can be calculated by extrapolating weak
coupling formulae, as we will see below).  We will argue that in a class
of M theory vacua the dominant boundary effect is Quantum
Chromodynamics, and a linear combination of the Kahler axions is a QCD axion.  A second phenomenological issue
arises from the fact that the unification scale is so close to $\lp$.
The most sensitive probe of such large
scales is proton decay.  Exact or approximate symmetries will
be essential in understanding why the proton is so stable.

A careful examination of the four dimensional
low energy effective theory gives rise to
other interesting observations.  The Kahler axion multiplet
 naturally gives rise to a no scale model with broken SUSY and vanishing
 cosmological constant as the leading term in a systematic computation
 of the effective potential.  A natural explanation of squark mass
 universality can also be obtained in this model.  The no scale
 structure and squark degeneracy are only valid in leading order in
 $(\Releven\Meleven)^{-1} \sim 0.1$.  It is unclear how far we can rely
 on these results as explanations of phenomena in the real world.

One of the most fundamental issues in string
theory is the question of how the moduli are stabilized.
In the weak coupling limit of string theory, moduli are
either exact or unstable.  If the theory describes nature,
one must hope that the moduli are stabilized at
a point in moduli space where semiclassical reasoning is not valid.
This raises the worry that one will not be able to predict anything from
string theory.   There will be no small parameter to explain
a small scale of supersymmetry breaking, for example, and the
smallness of the gauge couplings and their apparent unification
must be accidents. In ref. \banksdine, a solution to
this problem was suggested, exploiting the holomorphy
of the superpotential and gauge coupling functions,
and certain discrete symmetries.  It was
assumed that the compactification radii are of order
one in string units, and that string perturbation theory breaks down
even for small values of the dimensionless string coupling.  More precisely,
the model-independent dilaton $S$ was supposed large, while the other
moduli were of order one.  Stringy non-perturbative effects in
the gauge couplings and the superpotential were
shown to behave as powers of $e^{-S}$.  As a result,
one can understnad why supersymmetry breaking
is small and the gauge couplngs are unified, and
one predicts only tiny corrections to the lowest
order superpotential for matter fields.  Because
one is in a region where the dilaton superpotential
is monotonically decreasing, stabilization
of the moduli
must arise through large
corrections to the Kahler potential.  

The view that the compactification scale is large and
that the string coupling is very strong requires a reassessment
of this picture.  In the limit that the low energy, $11$-dimensional
supergravity description is valid, the theory suffers
{}from instabilities similar to those at weak coupling,
as we will see in some detail.  On the other hand, as we have
said, taking the $11$ dimensional parameters from the
``observed" four dimensional ones, nature would seem to
reside in precisely the regime where the long wavelength
description breaks down.  So it would seem reasonable to
hope that quantum $M$-theory effects are responsible
for the stabilization of the moduli.  One of the goals of the
present work is to explore this possibility, and to ask
what weak-coupling predictions, if any, survive into the
strong coupling regime.

In order to do this, it is necessary to understand as well
as possible the structure of the low energy theory in
the small $\epsilon$ regime.
One of the main results of ref. \wittenstrong\ is a computation
of the gauge couplings from an eleven dimensional perspective.
Studying the classical field equations, Witten finds that the
Calabi-Yau volume on the $E_8$ side decreases linearly
with $\Releven$, so that, for fixed $E_6$ coupling, the $E_8$ coupling blows up
at a finite value of $\Releven$.  In section 2.2, we point out that, exploiting
the holomorphy of the gauge coupling function, these functions
can be computed by weak coupling methods.   The imaginary
parts of the chiral
fields $\vec T$ and
$S$ (the usual moduli whose real parts describe the internal
radii and the four dimensional gauge coupling, respectively)
are the axions we have spoken of above.  They can be normalized
so that the theory is invariant under $2 \pi$ shifts of these
fields.  As a result, the gauge coupling functions are necessarily
of the form
\eqn\gaugecouplings{f^a={1 \over 32 \pi^2}(m^a S + \vec n^a \cdot \vec T)+
{\cal O}(e^{- (r S +\vec{s}\cdot \vec{T}) })}
where $m$, $\vec n$, $r$ and $\vec s$ represent sets of integers.
Perturbative heterotic string physics is valid when $S\gg 1; S\gg T$ and
$S/T^3 \gg 1$.  M theory is well
approximated by classical supergravity (SUGRA) 
when $S$ and $\vec T$ are both large
such that ${S\over \vert T\vert^3}$ is small.  However, we must avoid
regions where $S$ and $\vert T\vert$ are comparable and certain linear
combinations of them are small.  In these regions physics on one of the
boundaries of the eleven dimensional world is strongly coupled.
By holomorphy we can just as well calculate the linear terms in the
gauge kinetic functions at weak coupling.
Such weak coupling calculations have
been performed in the past for a variety of theories \ref\vadim{
V. Kaplunovsky and J. Louis, ``On Gauge Couplings in String
Theory," hep-th/9502077, Nucl. Phys.  {\bf B451} (1995) 53.} such as 
orbifold
models, and the difference of the $E_6$ and $E_8$ couplings
has been calculated for Calabi-Yau spaces.  In section $2.2$,
we point
out that the couplings themselves can be computed
directly for Calabi-Yau spaces, by dimensionally reducing
the Green-Schwarz counterterms introduced in ten-dimensions
to cancel anomalies.
\foot{The weak coupling calculation
which we will perform here -- and thus, in some
sense, Witten's eleven dimensional calculation,
has actually been performed some time ago
by L. Ibanez and P. Nilles, \ref\ibaneznilles{L.
Ibanez and P. Nilles, Phys. Lett. {\bf 180B} (1986) 354.}}
This gives the
couplings of the axions in the $T$ multiplets
to $F \tilde F$ and is easily supersymmetrized to
give the coupling of the full multiplet.  We find, as in ref.
\wittenstrong, that the sign of these couplings
is such that, for fixed $S$, the $E_8$ coupling
blows up at a finite value of the radius.

This fact is already quite striking.  It was probably ignored
in the past because it corresponds to a point of strong string
coupling.  As stressed in \wittenstrong, the blowing
up of the coupling may
have something to do with the stabilization of the
moduli.  As we will discuss in some detail,
in the weak coupling regime, one can determine
the potential for the moduli completely.
Gluino condensation gives rise to
a superpotential which behaves
as a power of $e^{- S + \vec{\alpha}\cdot \vec{T}}$.  In the semiclassical SUGRA regime, we will
fully determine the Kahler potential for the
moduli and matter fields.  Gluino condensation
then leads to a potential which
grows with radius for fixed coupling of the standard model gauge fields,
{\it i.e.} at weak coupling the dynamics tends to {\it shrink}
the eleventh dimension.  This is rather surprising, since
one might have expected that for widely separated
walls, the eleven-dimensional dynamics would become
free.  In the regime where M theoretical dynamics reduces to
supergravity, there is no way to prevent the shrinkage.  Thus,
quantum M theory is crucial to the stabilization of the radius of the 
eleventh dimension.
Similar remarks can be made about the size of X.  Phenomenology
indicates that it is quite close to $\lp$.  Consequently the
stabilization of this modulus probably also requires the intervention of
quantum M theory.

Given that the parameter $\epsilon$ is of order
$1$, it is not unreasonable to expect 
that quantum
$M$-theory dynamics stabilize the moduli at their observed
values.  Note that unlike the situation analyzed in \banksdine , there
is no mystery here about the weakness of the standard model gauge
couplings. The latter play no role in the stabilization of the moduli.
Nor do we need to invoke a premature breakdown of perturbation theory.
Weak coupling arises from geometrical factors of order one, primarily a 
factor of $2$ (which gets raised to the sixth power) between the linear
size of the Calabi Yau manifold on the $E_6$ boundary and $\lp$.
We will find, however, that residual constraints from $N=2$
supersymmetry  --
the no-scale structure we alluded to earlier --
 raise puzzles about how some of the moduli
are stabilized.

As in ref.  \banksdine\ we can
attempt to use holomorphy of the superpotential and gauge
coupling functions, together with exact discrete shift symmetries for the
moduli, to argue that certain semiclassical predictions
received only exponentially small corrections even at strong coupling.  
Now, however, the situation is more complicated.
We have noted above that in the M theory regime certain linear
combinations of the moduli can be small, and exponentials of these are
no longer suppressed.  By studying physics deep in the semiclassical
regime, where the Calabi Yau volumes are everywhere large, we will argue
below that these unsuppressed exponentials do not infect the predictions
of gauge coupling unification and ratios of Yukawa couplings. A crucial
ingredient of this argument is the use of holomorphy to extrapolate
semiclassical results into the regime of phenomenological relevance.

The rest of this paper is organized as follows.
In the next section we review the effective
theory in five dimensions which results from
compactification of $M$ theory on a Calabi-Yau
space.  We pay particular attention to
certain approximate symmetries which will survive in four
dimensions as Peccei-Quinn symmetries.  We then reduce the
theory to four  dimensions.  We give a weak coupling
string calculation of the dependence
of the coupling on the $11$-dimensional radius.
We discuss the form of the resulting Kahler potentials,
including restrictions inherited from the approximate
five-dimensional supersymmetry.  We point out that
there are several approximate Peccei-Quinn symmetries
which hold to an extremely high degree of
accuracy.  The axion associated
with one of these symmetries solves the strong
CP problem, but will violate the conventional cosmological bounds.
In section 3, we discuss the
problem of stabilizing the moduli, exhibiting the intriguing yet
rather problematic
no-scale structure.  We offer some speculations about how 
stabilization might occur and about the possible origin of
a hierarchy. Finally, in section 4, we discuss some phenomenological
and cosmological implications of these observations.

\newsec{Some Effective Field Theories}

\subsec{Effective Field Theory in Five Dimensions}

To begin, let us be more precise about the numerical values of various
parameters.  We do this not because of any illusion that the tree level
calculation of these parameters is immune to corrections, but in order
to orient ourselves.  The tree level fit to the fine structure constant
and the unification scale gives
\eqn\CYsize{R = 2\lp = (3 \times 10^{16}{}GeV)^{-1}.}
\eqn\orbsize{\Releven\Meleven = 8.}
Here, $L$ is the sixth root of the volume of X, $\Releven$ is the length of
the eleventh dimension ($\pi\rho$ in Witten's notation), $\Meleven=
\lp^{-1}$, and $\lp$ is the ninth root of $\kappa^2$, the coefficient of
the eleven dimensional Einstein action.  The fit of M theory to the real
world suggests that six of the dimensions are very small, one is one 
order of magnitude larger and the rest are at least as large as our
horizon volume.

We can also write a formula for the heterotic string tension in terms of
eleven dimensional quantities.  For large $\Releven$ we have an
approximate five dimensional SUSY, and this is a BPS formula which
receives no corrections.  Boundary effects and other breaking of SUSY
down to four dimensional $N = 1$, will give corrections to this formula
of order $(\Releven\Meleven)^{-1}$, which we will neglect. The string
tension formula can be obtained by the following
reasoning.  In ten dimensions, one has expressions for the
gauge and gravitational couplings in terms of the
tension\ref\hetii{D.J. Gross, J.A. Harvey, E. Martinec
and R. Rohm, Nucl. Phys. {\bf B267} (1986) 75}:
\eqn\tendkappa{\kappa_{10}^2 ={1 \over 4}\lambda^2
 (2 \alpha^{\prime})^4}
\eqn\tendg{g_{10}^2=\lambda^2(2 \alpha^{\prime})^3,}
where $\lambda$ is the dimensionless string coupling.
Comparing with the eleven dimensional expressions for these
quantities yields
\eqn\alphaprime{
 2 \alpha^{\prime}=
{(\kappa_{11})^{2/3} \over \pi \Releven ( 4\pi)^{2/3}}}
\eqn\numericalap{~~~~~~~~~~\approx {1\over 136} M_{11}^{-2}.}

Alternatively, we can use Polchinski's formula for the Dirichlet two
brane tension in Type IIA string theory\ref\rrd{J.Polchinski,
``Dirichlet Branes and Ramond Ramond Charges'', {\it Phys. Rev. Lett.}
{\bf 75}, (1995), 4724, hep-th/9510017.}\foot{We thank J. Polchinski for
guidance through the conventions of this paper.}, and the fact that the
heterotic string is just a two brane stretched between the walls of the
world.  We also need the Kaluza Klein relation between the ten and
eleven dimensional gravitational constants.  This calculation gives the
same result as above, if one is careful about factors of $2$ coming from
the relation between compactifications of M theory on a circle and an
orbifold.

Given the relatively large size
of $\Releven$, it is appropriate to
consider an effective five dimensional action for
physics at length scales larger than $\lp$ but smaller than $\Releven$.  We
will then reduce this to a four dimensional effective action for scales
longer than $\Releven$.  In the bulk, the five dimensional theory has
full five dimensional SUSY, and its lagrangian has been worked out by 
Antoniadis {\it et. al.}\ref\anton{I. Antoniadis,
S. Ferrara, T.R. Taylor, ```N=2
Heterotic Superstring and Its Dual
Theory in Five Dimensions,''
Nucl.Phys. {\bf B460} (1996) 489,
hep-th/9511108 }, following \ref\towns{M.Gunyadin, G.Sierra,
P.K.Townsend, Nucl. Phys. {\bf B242} (1984) 244;
Nucl. Phys. {\bf B253} (1985) 573.}.  
The volume of X is in a
hypermultiplet along with some of 
the purely internal and the dual of the purely
external components of the three form gauge potential.  The complex
structure moduli also pair up into hypermultiplets, with internal
components of the three form.  The quaternionic 
metric on this
space of hypermultiplets is not determined by general considerations.
For large volume it can be computed by Kaluza Klein technology.
However, equation \CYsize\ tells us that the volume is not large.
We expect M theory to give corrections to this metric of order $\lp
R^{-1}$. 

On the other hand, the volume preserving Kahler moduli,
are in vector multiplets, along with the integrals of the three form
over nontrivial $(1,1)$ cycles.  The bosonic part of the 
lagrangian for these multiplets is
given by\anton \foot{Micha Berkooz has pointed out to us that
the nontrivial background fields calculated by Witten, break $d=5$ SUSY.
Thus, there may be corrections to this lagrangian.  However, 
when $\Releven$
is much larger than $\lp$ the unknown dynamics of short distance M theory
should not be affected by this soft breaking of $d=5$ SUSY.  The
corrections should be calculable in low energy supergravity.  
That is, integrating out the unknown massive degrees of freedom
of quantum M theory, should give us a Lagrangian which is $d=5$
supersymmetric to leading order in $\Releven^{-1}$. The fields
which Witten calculates to have $N=2$ SUSY breaking VEVs are all in
hypermultiplets and they do not effect the vector multiplets to leading
order in the long distance expansion.}
\eqn\vectlag{{\cal L}_{vec} = G_{ab}[F_{\mu\nu}^a F^{\mu\nu b} + \partial_{\mu}X^a 
\partial_{\mu} X^b ] + C_{abc}\epsilon^{\mu\nu\lambda\kappa\sigma}
A_{\mu}^a F_{\nu\lambda}^b F_{\kappa\sigma}^c}
Here $G_{ab} = - \partial_a \partial_b {{\rm ln} N}$, where ${N} =
C_{abc} X^a X^b X^c$, with $C_{abc}$ the intersection numbers of the
corresponding $(1,1)$ forms.  The fields $X^a$ are constrained to
satisfy $N(X) = 1$.

This lagrangian is invariant under local gauge transformations of the
$h_{1,1}$ U(1) gauge fields which vanish at the boundaries of the fifth
dimension.  Now consider the transformation $\delta A_{5}^a = \partial_5
c^a$ with $c^a $ a linear function which vanishes only 
on the $E_8$ boundary.  
More microscopically, we view this transformation as originating from 
transformations of
the eleven dimensional three form gauge field by $\delta A_{i\bar{j}11}
= \partial_{11} d^a_{i\bar{j}}$.  Here we choose the gauge function so
that $\partial_{11} d^a_{i\bar{j}} = b^a_{i\bar{j}}$, with $b^a$ one of
the harmonic $(1,1)$ forms on the Calabi-Yau fiber at $x^{11}$, and so
that $d^a_{i\bar{j}}$ vanishes on the boundary. $c^a$ is the integral of
 $d^a$ over the $a$th $(1,1)$ cycle on the manifold (and we have renamed
the eleventh dimension the fifth).  This transformation is not a
symmetry of the system.  However, it is broken only by nonperturbative
physics which involves the $E_6$ boundary.  Loosely speaking,
nonperturbative effects on the $E_6$ boundary arise from
membranes stretched between the two boundaries, and Euclidean
5-branes wrapped around the Calabi-Yau manifold on this boundary.
These approximate symmetries become Peccei-Quinn symmetries of the
effective four dimensional theory.  We will estimate the dominant
symmetry breaking effects below.

\subsec{Four Dimensional Effective Field Theory}

We now want to reduce our resolving power and obtain a description of
the world on length scales longer than $\Releven$.  This will be an $N=1$
locally supersymmetric four dimensional field theory.  We first address
the question of the gauge couplings in this theory.  Witten has given us
an eleven dimensional calculation of the blowup of the $E_8$ coupling
when $\Releven$ reaches a critical value.  It is a remarkable example of the
{\it power of holomorphy}\ref\nati{N. Seiberg,
``The Power of Holomorphy,"
hep-th/9506077.} that this calculation can be
exactly reproduced by extrapolation of results for the weakly coupled
heterotic string.  

Witten determines the dependence of the $E_6$
and $E_8$ gauge couplings on the volume 
of the Calabi-Yau space and the radius
of the eleventh dimension.  However, if the
four dimensional effective coupling is small,
while the Calabi-Yau radius is large, it should
be possible to obtain this dependence from
a weak coupling computation.  The point is that
there is a regime of large radius (``$T$'') and small coupling
(large ``$S$''), such that the dimensionless
string coupling is small, and these couplings
can be computed in perturbation theory.  The gauge
coupling functions are holomorphic functions of $S$
and $T$.  They must also be invariant under
discrete shifts of $S$ and $T$.  With the normalizations
we will use, these shifts are,
\eqn\periodicity{S \rightarrow S +
2 \pi i ~~~~~ T \rightarrow T + 2 \pi i.}
As a result, up to terms which are exponentially
small for large $S$, the gauge couplings functions
$f_a$, must be given by
\eqn\gaugefunctions{f_a =m_a S + n_a T,}
where $m_a$ and $n_a$ are integers.  The
$m_a$'s are determined by the central
terms, $k_a$, in the Kac-Moody algebras.
The $n_a$'s can be obtained from a one
loop computation.

These couplings have been evaluated in the literature
for many special cases\ref\vadim{L. Dixon,
V. Kaplunovsky and J. Louis, Nucl. Phys.
{\bf B355} (1991) 649; J.P. Derendinger,
S. Ferrara, C. Kounnas and F. Zwirner,
Nucl. Phys. {\bf B372} (1992) 145; I.
Antoniadis, K. Narain and T. Taylor,
Phys. Lett. {\bf B267} (1991) 37; V. Kaplunovsky and J.
Louis, ``On Gauge Couplings
In String Theory," Nucl. Phys. {\bf B444} (1995) 191,
hep-th/95092077}.
For large radius Calabi-Yau compactifications,  a formula
has been presented for the difference of the
$E_6$ and $E_8$ couplings\vadim.
However, for large radius, the separate
couplings are well defined and it is actually a simple matter to determine
them.  The point is that for large
radius, these couplings can be obtained by
reduction of the ten-dimensional effective action.
In particular, in terms of component fields,
these couplings imply couplings of certain
``axion-like'' fields to $F \tilde F$.
These axions correspond to particular excitations
of the antisymmetric tensor field, $B_{MN}$, with
indices in the internal space.  Such couplings are
necessarily linear in $B$ and involve products
of $F_{\mu \nu}$, i.e. from a ten-dimensional
perspective they are precisely the terms which
appear in the Green-Schwarz counterterms.
So it is only necessary to reduce the Green-Schwarz
counterterms to four dimensions.

Before examining the Green-Schwarz counterterms themselves,
a few preliminaries are necessary.  First, we must
determine the excitations of the $B$ field
corresponding to the various axions, and how they
fit into chiral multiplets.  The necessary expressions
appear in ref. \ref\dsww{M. Dine, N. Seiberg, X.-G. Wen
and E. Witten, Nucl. Phys. {\bf B289} (1987) 319,
{\it Nucl. Phys.} {\bf B278} , (1986),769.}
The axions are in one to one correspondence with
harmonic $(1,1)$ forms, $b_{i,\bar i}^{(a)}$.
These are conventionally normalized so that
\eqn\bnorm{\int_{\Sigma_a}b^{(b)} = \delta^b_a}
where $\Sigma_a$ are a basis of nontrivial
closed two-dimensional sub-manifolds.
In terms of these, and adopting units
with
$2 \alpha^{\prime}=1$,
the action takes the form
\eqn\action{I={-i \over 2 \pi} \int d^2 z \sum
(2 \pi)[(r_a+i \theta_a)b_{i \bar i}^{(a)}
\bar \partial X^i \partial X^{\bar i}
+((r_a-i \theta_a) b_{i \bar i}^{(a)}
\bar \partial X^{\bar i} \partial X^{i}].}
By virtue of the normalization of the
$b^{(a)}$'s, the coefficients
of the $\theta_a$'s are quantized,
and $\theta_a$ has period $2 \pi$.
As we will now show, $\theta_a$
is the imaginary part of the chiral
field whose real part is $r_a$.
Note that $2 \pi r_a$ is what one
would call the radius-squared of the
internal space.

In order to determine the structure of the
four dimensional chiral fields, it is
necessary to adopt some conventions.
We take the ten-dimensional fields
to satisfy $\Gamma_{11}=1$, where
$\Gamma_{11}= \Gamma_1 \dots \Gamma_{10}$.
In making the reduction to four dimensions,
we introduce three complex coordinates,
$x_i$ and $\bar x_i$ (this was implicit
in the discussion above), and a corresponding
set of $\gamma$ matrices.
In particular, if we define
\eqn\complexcoord{X^1 = x^1 + i x^2 ~~~~~X^{\bar 1}= x^1 -i x^2 }
etc., and if we define corresponding six dimensional
$\gamma$ matrices, $d^i$ and $d^{\bar i}$, then
we can define ``states'' by
\eqn\gammaspace{\vert 0 > ~~~~~\vert \bar i \rangle =
d^{\bar i} \vert 0 \rangle ~~~~~\vert k \rangle =
d^{\bar i} d^{\bar j}
\vert 0 \rangle ~~~~~ \vert \bar 0 \rangle
= d^{\bar 1} d^{\bar 2} d^{\bar 3} \vert 0 \rangle.}
Calling
$d^7 = -i d^1 \dots d^6$, $\vert 0 \rangle$ has $d^7$=1,
and the chiralities of the other states follow immediately.
In particular, the states $\vert i \rangle$ have chirality
one both internally and in four dimensions.
So vertex operators of the form
\eqn\twentyseven{V_{27} = b_{i \bar i} \lambda^{\bar i} DX^i}
are vertex operators for $27$'s with positive chirality.
Note, however, that when trying to identify these operators
with {\it fields}, it must be remembered that the
vertex operators are like creation operators, i.e. they
are like complex conjugates of fields.  Similarly,
we can read off the operators for the
moduli, from eqn. \action.
In particular, the chirality plus field is the one which
multiplies $DX^i$, but complex conjugated as described above, i.e.
$r_n+i \theta_n$

Now we can turn to the Green-Schwarz term.  This term has been
evaluated in various places.  We choose to take the result
{}from ref. \ref\lerche{W. Lerche, B.E.W. Nilsson, and A.N.
Schellekens, Nucl. Phys. {\bf B289} (1987) 609.}:

\eqn\gs{{1 \over 24\times 12} {1 \over (2 \pi)^5}
\int B [Tr F^4 -{1 \over 300}(Tr F^2)^2
-{1 \over 10} Tr F^2 tr R^2 + 3Tr R^4+{3 \over 4}
tr (R^2)^2]}
We can dimensionally reduce this immediately.  Break up
$F$ into parts with indices in four dimensions
and indices in the internal six dimensions.  Replace
$B$ by $2 \pi \theta_a b^{(a)}$.  Recall
that $\rm Tr(F^4)={1 \over 100} (\rm Tr F^2)^2$,
and $\rm Tr F^2 = 30 \rm tr F^2.$  One then obtains,
for the $E_8$ coupling to the axion,
\eqn\couplings{\theta_a {1 \over 32 \pi^2}\int d^4 x F \tilde F
\int {b^{(a)}\wedge F \wedge F \over 8 \pi^2}.}
For the $E_6$ coupling, one obtains the same result
but with the opposite sign.

In order to finally determine the sign of the coupling
of the modulus to the gauge fields, one notes that
that the sign of the coupling of the imaginary part
to $F \tilde F$ is opposite to that of the coupling
of the chiral field to $W_{\alpha}^2$\ref\wb{J. Wess
and J. Bagger, {\it Supersymmetry and Supergravity},
Princeton University Press, Princeton (1983).}.
So we see that the $E_8$ fields couple to
$S-T \int {b \wedge F \wedge F \over 8 \pi^2}$
while the $E_6$ fields couple to the same
combination but with the opposite sign for the
$T$ term.

Finally, we can compare this with Witten's result.  Using
the formula for $\alpha^{\prime}$, eqn. \alphaprime,
and Witten's expression for the difference of the
$E_8$ and $E_6$ couplings,
\eqn\wittencoupling{\delta \alpha^{-1} =
{2 \over (4 \pi)^{4/3} \kappa^{2/3}}2 \pi^2 \Releven
\int {1 \over 8 \pi^2}\omega \wedge (F \wedge F -{1 \over 2} R \wedge R)}
we have, in units with $2 \alpha^{\prime}=1$
\eqn\fourdversion{={1 \over 8 \pi^2} \int {\omega \wedge F \wedge F \over 8 \pi^2}}
where we have taken the spin connection to equal the
gauge connection.
To obtain the corresponding term in the action involving $F\tilde F$,
one needs to multiply this expression by
$1 \over 16 \pi$.  
Again, the properly normalized fluctuation of $B$
($\omega$) contains a factor of $2 \pi$, so the difference of the
two couplings is the same as expected from eqn. 2.12.  However,
it is also clear that we cannot identify the volume on the
$E_6$ side with the weak coupling $S$; it would appear to
be something like $S-cT$.

In order to be
more precise about the comparison between the weak coupling
results and Witten's, we must pay more attention to the proper
definition of four dimensional chiral superfields in terms of higher
dimensional geometry.  In the weakly coupled region, there is only one
Calabi Yau volume, while in the M theory regime, we must specify
precisely what average over the fifth dimension we are using.  Thus, as
we have seen above, it is wrong to identify the real part of the 
chiral superfield $S$ of the weakly
coupled heterotic string with the volume of the Calabi Yau manifold on
the $E_6 $ boundary, which Witten uses to parameterize his results.  The
weak coupling calculation shows that the $E_6$ coupling is a linear
combination of $S$ and the $T^a$, and it is this linear combination
which is identified with the volume on the $E_6$ boundary.  
We have found it useful to pass through $5$ dimensions in our search for
a good parameterization of the space of chiral superfields in the four
dimensional effective field theory.  In particular, we want to keep
track of the $h_{(1,1)}$ approximate $U(1)$ symmetries which are
unbroken by strong $E_8$ dynamics.  These act on chiral superfields
which are defined in terms of functions with boundary conditions on the
$E_8$ boundary.  That is, shifts of the imaginary part of these
superfields are would be five dimensional gauge transformations with
gauge function defined to vanish on the $E_8$ boundary.  We will call
the $h_{1,1}$ axion chiral multiplets $Y^a$.
It is convenient then to parameterize the volume and complex structure
moduli by their values on the $E_8$ boundary as well.  As noted in a
previous section, these
belong to five dimensional hypermultiplets.  However, only one chiral
field from each hypermultiplet survives the breaking of SUSY that
accompanies the reduction to four dimensions.  We will denote the
superfield whose real part is proportional to the volume of $X$ on the
$E_8$ boundary by ${\cal S}$.  The normalization is fixed so that shifts
of the imaginary part of ${\cal S}$ by $2\pi$ are exact symmetries of
the theory.  The complex structure moduli on the $E_8$ boundary are
denoted by $C_{\alpha}$.  Note that although all of these fields are
defined in terms of boundary conditions, they are what we will later
describe as bulk moduli.  The
boundary conditions determine the behavior of the classical vacuum
configuration throughout the fifth dimension.  The action for making a
small spacetime dependent deformation of these boundary conditions will
be proportional to $\Releven$.  

With these definitions we can write our weak coupling results for the
gauge kinetic functions in terms of the fields ${\cal S}$ and $Y^a$ in
the M theory regime.  We have ${\cal S} = 
S-T^a \int {b_a \wedge F \wedge F \over 8 \pi^2}$.

To summarize, the basic phenomenon observed by Witten, {\it i.e.} that
gauge couplings can blow up in the region of moduli space where the
Calabi Yau volume is larger than the string scale,
is evident in extant weak coupling calculations.  It has probably been
ignored in the past because it only occurs when the heterotic string is
strongly coupled, but analyticity and discrete symmetries allow us to
reliably compute in this region.  The perturbative computation
reproduces the fact that the term in the $E_8$ coupling function linear
in the moduli vanishes (and thus the gauge coupling becomes strong)
at a point in the M theory regime.  In addition
it enables us to identify the weakly coupled moduli fields as particular
linear combinations of the fields ${\cal S}$ and $Y^a$ which have simple
properties in the M theory region.
 Once the $E_8$ coupling becomes strong
however, we can no longer neglect possible exponential terms in the
gauge coupling function.  In the low energy $E_8$ gauge theory, an
accidental $U(1)$ symmetry prevents the occurrence of such terms, but in
M theory we do not expect to have such a symmetry.  Thus, although we
know that the coupling becomes strong, we do not know that it becomes
infinitely strong.  In the strong coupling region we do not have a
reliable calculation either of the $E_8$ coupling itself, or of the
superpotential for the moduli which is generated by the strongly coupled
dynamics. 

One may worry that similar incalculable effects will infect the
computation of the coupling functions on the $E_6$ boundary.  This could
completely ruin predictions of coupling unification.
We know of
no symmetry argument which rules this out, but we believe that the
following physical argument is plausible.  Let us examine the region
where the Calabi Yau volume is much larger that $\lp^6$.  In this
regime, the $E_8$ coupling becomes strong only at very large $\Releven$.
Thus, there is a regime in which $\Releven$ is large and the $E_8$ 
coupling is still small enough that nonperturbative dynamics is well
approximated by a very dilute gas of small instantons.  In addition, 
since the instanton density is exponential in the coupling, the
average instanton spacing can be taken much larger than $\Releven$.  
In this limit, the dominant effect on the lagrangian of the
$E_6$ boundary will come
{}from the local influence of single instantons. $E_8$
instantons are $5$ branes in eleven dimensional space.  Their effect on 
the $E_6$ boundary must fall like $R_{11}^{-3}$ as $\Releven$ is increased.
Thus they cannot give rise to effects on the $E_6$ coupling functions
which grow exponentially with $R_{11}$.  Indeed, Green's functions made up
of fields which live purely on the $E_6$ boundary cannot soak up the
$E_8$ instanton zero modes, and get no contribution from these
nonperturbative configurations.  We have made this argument for very
large $V$ and $\Releven$, but holomorphy tells us that if the growing
exponentials are not present in this regime, they are not present at
all.  Coupling unification {\it is} a prediction in the M theory region
of moduli space.

One advantage of our weak coupling calculation of vacuum polarization
functions is that we can easily extend it to the case where 
$E_8$ is broken by Wilson lines.  In fact, it is not difficult
to see that the result is unchanged in the presence of Wilson lines.
At large radius, on the torus, one must compute an expectation
value of the form
\eqn\wilsonlines{\langle V_B V_A V_A V_A V_A \rangle}
where $V_B$ is a vertex operator for the antisymmetric
tensor, and $V_A$ is a vertex operator for a gauge field.
One can take, say, $V_B$ in the $-1$ superconformal ghost
number picture, and the $V_A$'s in the zero ghost picture.
As in the flat space calculation, the term with an $\epsilon$
tensor arises from the sector with $(P,P)$ boundary
conditions for the right movers.  In the large $R$ limit,
there is an (approximate) zero mode for each of the
$\psi_I$'s.  This is just the correct number of
zero modes to be soaked up by the five vertex operators
in eqn. \wilsonlines.  The momentum factors in the
four gauge boson vertex operators then give $F^4$.
Because the fundamental group of the non simply connected
Calabi Yau manifold acts freely on its covering space, at large
radius one just has an ordinary momentum integral
to do, up to terms which are down by powers of $1/R$.
Such terms have the wrong $R$ dependence to correct
the modulus-dependence of the gauge couplings.

The rest of the calculation is as in ten dimensions.  The
right moving boson and fermionic contributions cancel.
Level matching then implies that only states with $\bar L_0 = 0$
contribute on the left.  This is identical to the situation
without the Wilson line.

\subsec{Kahler Potentials}

The dynamics of SUSY breaking in the M theory regime is, as usual in
string theory, intimately connected with the stabilization of the
moduli.  In the M theory regime, the moduli break up into several
distinct classes.  All moduli originate as dimensionless deformations of
a supersymmetric classical ground state of a theory with fundamental
mass scale $M_{11}$.  However, fields that originate in the bulk of
eleven dimensional spacetime, have kinetic terms in the effective four
dimensional theory which are proportional to $\Releven$.  In particular this
is the case for the four dimensional metric and this is part of Witten's
proposal for the origin of the large ratio between the four dimensional
Planck mass $\Mfour$ and the eleven dimensional Planck scale $\Meleven$.
Thus, we should imagine that in the conformal frame fixed by 11
dimensional SUGRA the Kahler potential for all of the bulk
moduli\foot{This is the term which we use to describe chiral superfields
which originate as modes of bulk fields in five dimensions.}
 has a coefficient of order ${M_4^2 \over 8\pi}\equiv m_4^2$.  
When we rescale these
 fields, $B_i$, to give them their proper dimension, their lagrangian
will be a function of ${B_i \over m_4}$.  Note that it is the reduced
Planck mass $m_4$ that we choose in this formula rather than the Planck
mass itself.  Historically, it has been natural to associate the mass
associated with Newton's constant as the fundamental mass scale of
quantum gravity.  However, in the M theory regime at least, it 
is a low energy artifact.  $m_4$ is the parameter which appears in all
formulae in the M theory regime.

In writing a supersymmetric four dimensional lagrangian, it is
convenient to choose a conformal frame in which the Einstein term does
not depend on the chiral superfields.  This is the frame in which
textbook expressions for the supergravity potential are written.
In this frame, the coefficient of the Einstein term and of the Kahler
potential for dimensionless bulk moduli fields is $\Meleven^2$.  We will
refer to this as the canonical frame.  Note that this is different from
the Einstein frame, where the coefficient of the four dimensional
Einstein lagrangian is $m_4^2$.  This is a consequence of the fact that
$\Meleven$ is the fundamental scale, while $m_4$ is a function of the moduli.

Among the bulk moduli will be those that descend from components of
vector multiplets in 5 dimensions.  $h_{1,1}$ of these can be
associated with Kahler deformations of the Calabi Yau
manifold.  The way in which these emerge from the 5 dimensional
lagrangian has been described in \towns 
.  Remember that the five
dimensional theory contained $h_{1,1} - 1$ vector multiplets, whose scalar
components live on a manifold with coordinates $X^a$ satisfying the constraint
$N(X) = 1$.  The dimensionally reduced theory is conveniently described
in terms of the complex fields $Y^a = R_{11} X^a + i A_5^a$ which are
the scalar components of chiral superfields.  These fields are
unconstrained and have (in canonical conformal frame) the
Kahler potential $- {\rm ln}\  N(Re Y^a )$.  In this approximation, 
the theory is invariant under
continuous shifts of the imaginary parts of the $Y^a$.  In the quantum
theory, we expect this to be broken to a discrete shift symmetry.
However we have argued above that the symmetry breaking is entirely due
to stretched membranes and to fivebranes embedded in the $E_6$ boundary.

We can estimate the size of the stretched membrane contribution in two
ways.  First a naive eleven dimensional calculation suggests a
PQ symmetry breaking term of order
$e^{- c \Meleven^3 R^2 \Releven} $
This is the same form as the PQ breaking term
which arises from a single worldsheet instanton in the weak
coupling theory.  We can
in fact,
reproduce this result by analytically continuing the world sheet
instanton contribution of weakly coupled string theory.  This has the
form $e^{- c R^2}$ in units with $2 \alpha^{\prime}=1$.
Inserting the formula for
the string tension in terms of eleven dimensional quantities we
get $e^{- c 4^{2/3} \pi^{5/3}\Meleven^3 R^2 \Releven}$.  
The latter derivation allows us to compute the
precise coefficient, $c$, in the exponent for specific Calabi Yau manifolds.
It also leads us to another example in which symmetry and
holomorphy
arguments are enhanced by an appeal to physical intuition.
Symmetry and holomorphy would allow us to add a term to the
space time
superpotential which breaks the axion shift symmetries and vanishes
only
when the $E_8$ coupling is weak.  This could give the axions a large
mass when the $E_8$ coupling is strong.
The physical picture of stretched membranes assures us that this
does
not occur.  Strong $E_8$ coupling might modify the contribution of a
membrane instanton on its boundary.  This should appear as a
multiplicative factor of order one 
in the instanton amplitude, and will not change our
estimate of its order of magnitude.

To estimate the value of the axion mass, 
we plug in the values of $R$ and $\Releven$ from our fit.  These are
determined in terms of $\Meleven$ so we must also use
the expression, eqn. \alphaprime,
for $\alpha^{\prime}$

The axion mass vanishes in the limit of supersymmetry breaking;
it is thus expected to be of order the SUSY breaking scale to the
fourth power.  Assuming that this scale is of order $10^{11}$ GeV,
yields an axion potential of order
\eqn\axionmass{V_a=e^{- 544 c} 10^{44}\ 
{\rm GeV} \sim 10^{(44 - 236 c)} \rm GeV.} 
This should be
compared with the Quantum Chromodynamic
 contribution to the axion potential
which is $\sim 10^{-4}$ in GeV unitsÏ.  For $c > 0.1$, the QCD contribution
dominates, and the model will solve the strong CP
problem.    In orbifold examples, $c \sim (2\pi)^2$ and
the stretched membrane contribution is completely negligible.
It appears then to be a general feature of the M theory region of moduli
space that there are $h_{1,1}$ axion fields which get their mass mainly
{}from nonperturbative effects on the $E_6$ boundary.  The strongest such
effect, if the gauge group is broken to the standard model, is QCD, and
one of the axions will solve the strong CP problem. 
Others will get their mass only from weak instantons, and from
stretched membranes.
These very light axions will have Compton wavelengths of 
astrophysical magnitudes.  However, their coherent couplings to matter may be
suppressed relative to gravity by as much as low energy CP violation.
In this case we believe that they may be compatible with observation.
If not, M theory will only describe the real world if $h_{1,1} = 1$
\foot{However, see the comments about boundary axions below.}.
In any event, in the M theory region of moduli space, axions solve the
strong CP problem.
The relevant invisible axion violates the cosmological axion
bound.  We will comment on this in the cosmology section below.

The low energy spectrum also includes fields which originate as modes on
the boundary of the five dimensional world.  Apart from the gauge
fields, there are the moduli of the $E_8$ gauge bundle on the $E_6$
boundary, and quark, lepton and Higgs superfields, as well as possible
exotic matter.  We denote the generic chiral multiplet originating on
the boundary as an edge field, $E_I$.
The Kahler potential for these fields is of order
$\Meleven^2$, so that when they are made dimensionful, their lagrangian will
depend on ${E\over \Meleven}$.  In general, it will depend on the bulk
moduli as well, and will be a correction to the Kahler potential of
these fields.  When $R_{11}$ and $R$ are large, but
$\epsilon$ is small, it is a simple matter to determine
the Kahler potential for these edge states.  It is, in fact,
precisely the same as on the weakly coupled string side.
To see this, one simply has to consider the lagrangian
for the edge states, which for the bosonic fields takes the form:
\eqn\edgel{{\cal L}_{e}= -{1 \over 8 \pi (4 \pi \kappa^2)^{2/3}}
\int d^{10} x \sqrt g \tr F^2.}
Reducing this lagrangian is similar to reducing the
usual ten-dimensional supergravity lagrangian on a
Calabi-Yau space.  The factors of $R_{11}$ work
out correctly.  In particular, if one first reduces to ten-dimensions,
it is necessary to rescale the ten-dimensional metric
by $g_{MN} \rightarrow \Releven^{-1/4} g_{MN}$.
This gives $\Releven^{-3/4}$ in front of the gauge
term, which is the conventional form of the ten-dimensional
action.

It is curious that the Kahler potentials for all
of the fields have the same form at both extremely weak and
extremely strong string couplings.  It is not
clear that this is enforced by any symmetry.  Moreover,
we have seen that the identification of the fields
$S$ and $\vec T$ is different in the two regimes.
Nevertheless, perhaps it holds some deeper meaning. 

Finally, let us note that the boundary moduli may provide us with
another candidate for the invisible axion.  Indeed, in \dsww,
it was shown that many $(2,0)$ moduli
might receive masses of order $e^{- T_h R^2}$.
This is a superpotential
calculation, and may be analytically extrapolated into the M theory
regime.  Since it refers to fields which live on the $E_6$ boundary, it
will not be affected by strong coupling dynamics on \lq\lq the other
side of the world\rq\rq .  If these $(2,0)$ moduli affect the $E_6$
gauge couplings at one loop in heterotic perturbation theory, as is
almost certainly the case, then they will provide another contribution
to the QCD axion.  The true axion will be a linear combination of these,
and the $h_{1,1}$ moduli discussed above.  However, because the boundary
moduli have decay constants of order $\Meleven$ rather than $m_4$, the
dominant component will be a boundary modulus.  This will ameliorate the 
cosmological axion problem.

\newsec{Mechanisms for Stabilizing the Moduli}

In the limit that the classical eleven dimensional description is good,
we expect to find the usual problem of runaway in the various
moduli.  If we simply consider compactification
with gauge group $E_6 \times E_8$, we can compute the
potential due to gluino condensation of the ``far side."
We do not need to think carefully about the interactions between
the two walls to do this, since we have already determined
the four dimensional Kahler potential, and the superpotential
due to gluino condensation follows, as in ref. \ref\gluinocondensate{
J.P. Derendinger,
L.E. Ibanez and H.P. Nilles, Phys. Lett.
{\bf 155B} (1985) 65; M. Dine, R. Rohm,
N. Seiberg and E. Witten,
Phys. Lett. {156B} (1985) 55.} from symmetry considerations.
One obtains, then, a potential identical to that at weak coupling.
It tends to zero as
\eqn\vasymptotic{V \approx \vert Y\vert^{-3} e^{-  {\cal S}/b_o}}
This potential favors large Calabi-Yau volume 
on the $E_8$ boundary and large $\Releven$.  This is a region where the
supergravity analysis should be completely valid,
so we have encountered the eleven dimensional version
of the stability problem.
Perhaps, however, the fact that, for fixed $E_6$ gauge coupling, the
potential forces $\Releven$ to zero is a hopeful sign.
This follows from the fact that for fixed $E_6$ coupling, the $E_8$
coupling (and thus the strength of the gaugino condensate) decreases
with $\Releven$. 

We turn, then, to a discussion of what sorts of physics might
stabilize the moduli.
We begin by
discussing the dynamics of the strongly coupled gauge theory on the
$E_8$ boundary.
 The proximity of the phenomenologically determined value of $\Releven$
to the strong coupling point motivates us to search for a mechanism
involving  the strong gauge dynamics which freezes some of the fields.  

As we argued in the previous section, the fields associated with five
dimensional vector multiplets do not participate in the strong dynamics.
The superpotential generated by $E_8$ and other quantum effects in M
theory will be a function of ${\cal S}$ and perhaps of the complex
structure moduli, but will not depend on the fields $Y^a$.  
Label the fields on which it does depend $Z^A$.  Then the potential will
have the form
\eqn\pot{V = M_{11}^4 {e^{K} \over N}(K^{AB}F_{A}\bar{F_B} + [G^{ab}G_a G_b -
3]\vert W\vert^2 )}
Here $G\equiv -{\rm ln} N$ and $G_a , G_{ab}, etc.$ refer to derivatives
with respect to $Re Y^a$ ($G^{ab}$ is the inverse metric).  This
expression is the first term in an asymptotic expansion of the potential
for large $Y^a$.  The equations $F_A = 0$ have a solution at ${\cal S} =
\infty$, the weak coupling region referred to above.  Generically, we
may expect them to have a solution for finite values of ${\cal S}$ as
well.  When ${\cal S}$ is small, the theory is strongly coupled and the
Calabi Yau volumes everywhere small (at finite $\Releven$), and we
can calculate neither the superpotential nor the Kahler potential.  It
is reasonable to postulate the existence of a discrete set of solutions
to these $k$ equations for $k$ complex unknowns.  Furthermore,
generically, $W$ will not vanish at these points.  In regions where
${\cal S}$ is relatively large it may be a good approximation to neglect
higher order terms in the superpotential, while retaining the
complicated Kahler potential.  The leading term in the superpotential
has the form $e^{-{{\cal S}\over b_0}}$ where $b_0$ is the first
coefficient in the renormalization group beta function.  The corrections
are powers of $e^{-{\cal S}}$ multiplied by the leading term or by $1$.

 We now note the
remarkable property\ref\noscale{E.Cremmer, S. Ferrara,
C.Kounnas, D.V.Nanopoulos,
{\it Phys. Lett.} {\bf 133B}, (1983), 61;J. Ellis,
A.B. Lahanas, D.V. Nanopoulos
and K. Tamvakis, Phys. Lett.
{\bf 134B} (1984) 429;
J. Ellis, C. Kounnas and D.V. Nanopoulos,
Nucl. Phys. {\bf B241} (1984) 406;
Phys. Lett. {\bf 143B} (1984) 410; J. Ellis,
K. Enqvist and D.V. Nanopoulos,
Phys. Lett. {\bf 147B} (1984) 99.}
of the Kahler potentials for the axion
multiplets, which has been widely exploited in {\it no scale} models:
{\it the term in square brackets in \pot\ vanishes identically for any
$W$ and any value of $Y^a$}.  As a consequence, the submanifold with
$F_A = 0$ of the full moduli space is, in the current approximation, a
stationary manifold of the potential, with broken supersymmetry and
vanishing cosmological constant.  Moreover, the scale of SUSY breaking
is as yet undetermined, since it depends on the values of the $Y^a$.

The $Y^a$ will be determined by terms higher order in the $Y$ expansion
of the Kahler potential.  At order ${1\over \vert Y\vert}$ we also
encounter terms in the Kahler potential that involve the boundary
fields.  These include quarks,$ Q^i$, and moduli of the gauge bundle that
breaks $E_8$ to $E_6$.  To this order, the Kahler potential will have
the form 
\eqn\Kahler{G = - {\rm ln} \ N(Re Y) + h(Y,E) + h_{ij}(Y) {Q^i}^* Q^j}
where $h$ and $h_{ij}$ are homogeneous of degree minus one in $Y^a$.
They can also depend on the gauge bundle moduli, $E_I$.  
We will assume that there is a solution of the equations 
\eqn\edgesusy{{\partial h(Y,E)\over \partial E_I} = 0,}
which fixes the value of the gauge
bundle moduli.  With this assumption, there is only one term of order
$Y^{-1}$ and quadratic in quarks which appears to depend
 on a matrix other than 
$h_{ij}$.  It is proportional to
\eqn\badterm{L_a L^{ab} h_{ij,a}(Q^i )^* Q^j}
where $L \equiv {\rm ln} N$.  $N$ is a homogeneous polynomial, so 
$L_a = - L_{ab} Y^b$ and the dangerous term is proportional to $Y^a
h_{ij,a} $.  To leading order in $Y^{-1}$, this is proportional to
$h_{ij} $ itself.  Thus the squark mass matrix is proportional to the
matrix in the quark kinetic term and we have universality.  Corrections
to this will be of relative 
order ${1\over \vert Y\vert} \sim 10^{-1}$ 
(here we use the phenomenological fit to the value
of $\vert Y\vert \sim \Releven$ since we are not yet able to calculate
it theoretically).  

The value of the $Y^a$ will be determined by minimizing 
the potential with $Q^i = 0$ and $E^I$ determined by equation \edgesusy
.  This procedure will have the usual philosophical problem discussed in
\dineseiberg .  Minimization is achieved only by balancing terms of
different orders in $Y$, even though $Y$ is large.  There are several
differences from the analogous problem in weakly coupled string theory.
There one is forced to contemplate cancellations between different
exponentials of a large number.  Here we have a Laurent series in $Y^a$,
and $\vert Y\vert$ must be of order $10$ in order to explain
the ratio between the unification scale and the Planck scale.  
A second contrast with the weakly coupled problem is that
we seem to have solved at least one of the
stability problems of the weakly coupled theory.  ${\cal S}$ is
presumed to be fixed in the {\it strong} coupling region by the equation
$F_S = 0$. Note that this is completely compatible with the fact that
the gauge theory on the $E_6$ boundary is weakly coupled at the
unification scale. 

Unfortunately, this argument leaves us with a puzzle about the scale of
SUSY breaking.  In the true strong coupling regime, the superpotential
generated by nonperturbative dynamics on the $E_8$ boundary is
of order $\Meleven^3$.  The gravitino mass is then fixed to be of order
$\vert {W\over \Meleven^3}\vert \vert Y\vert^{- 1} \Meleven 
\sim 10^{15}$ GeV.  In order to get
the right scale of SUSY breaking, we must assume that the superpotential
generated by the strong $E_8$ dynamics is of order $10^{-12}$ in eleven
dimensional units.  This suggests that the coupling is not terribly
strong and very probably that the gauge group is smaller than $E_8$.
For a gauge group $G$ with $k$ instanton zero modes in the adjoint
representation, the implied $G$ fine structure constant at the unification
scale is ${0.45 \over k}$.  

Another problem, which loses none of its severity through familiarity, is
that we do not have an explanation of the value of the cosmological
constant.  The no scale cancellation actually works through order $\vert
Y\vert^{-1}$ but fails at higher order.  
The fact that the vacuum energy density will be smaller by a factor
of $100$ than in a typical hidden sector model with the same value of
the gravitino mass is perhaps suggestive,
but hardly represents a solution
of the cosmological constant problem.

To conclude the discussion of this scenario, we briefly note the
properties of the moduli.  The bulk moduli coming from ${\cal S}$ and
the complex structure of $X$ will have masses of order the gravitino
mass.  Their kinetic terms are of the same order as the Einstein term in
the action, so their couplings to ordinary matter will be suppressed by
powers of $m_4$. In Einstein frame (the frame in which the coefficient
in front of the Einstein lagrangian is $m_4^2$) they will have potential energies of
order $ m_{{3\over 2}}^2 m_4^2$.
The boundary moduli $E_I$ have potentials of order $\vert
Y\vert^{-1} m_{{3\over 2}}^2 \Meleven^2$ in the canonical frame lagrangian.  
However, in this frame their kinetic terms also carry an inverse power
of $\vert Y\vert$.  In Einstein frame this means that they have
potentials of the form $ m_{{3\over 2}}^2 \Meleven^2 V({E\over
\Meleven})$.  Thus their masses are of order $m_{{3\over 2}}$ and their
couplings to matter are inversely proportional to $\Meleven$.  

The bulk moduli associated with the real parts of the axion multiplets
have a potential which is suppressed by two powers of $\vert Y\vert$
relative to the other bulk moduli.  Thus, their mass is of order
$10^{-1} m_{{3\over 2}}$ or $100$ GeV.  Their couplings to matter are
nonrenormalizable and scale with $m_4$. The QCD axion has a mass of
order $10^{-10}$ eV and decay constant of order $m_4$.  Its {\it
coherent} couplings to matter are further suppressed by the same factors
that suppress any low energy CP violation.  If $h_{(1,1)} > 1$ there
will be more of these multiplets.  Now however the axions will be
extremely light as noted above. 

If there are also boundary contributions to the QCD axion
then the true axion decay constant will be $\Meleven$.  We will also
have a definite prediction of a very light axion which would contribute
to long range spin dependent forces and very weak (compared to gravity)
long range coherent forces.  

\newsec{Low Energy Constraints on Planck Scale Physics}

The replacement of the Planck scale $M_4$ by $\Meleven$ as the threshold
for
as yet incalculable quantum gravitational effects sharpens the
constraints on physics at ordinary scales from possible
higher dimension operators.

The most important such effect is the lowering of the scale of
dimension five baryon number violating operators by two or three orders
of magnitude.  Discrete symmetries which eliminate
or suppress dimension five
operators become absolutely imperative.  The constraint from
gravitational physics is now of the same order as that conventionally
quoted for grand unified models. 
In a similar manner, we find a new estimate for gravitational
contributions to neutrino masses.  

We also find a stronger constraint on models which invoke
pseudogoldstone bosons of accidental continuous symmetries.  Previously,
one argued that a renormalizable theory at scale $f$ might spontaneously
break an
accidental continuous symmetry, producing a Goldstone boson with decay
constant $f$.  If gravitational physics breaks all global symmetries
(this is certainly the case in string theory) we expect a Goldstone
boson mass to be generated.  It will be of order ${f^{\ha (d-4) + 1} \over
M_G^{{\ha (d-4)}}}$
 where $M_G$ is the scale of gravitational effects and $d$ is
the dimension of the leading operator which breaks the
symmetry\ref\jmr{M. Kamionkowski
and J. March-Russell,
Phys. Lett. {\bf 282B} (1992) 137;
R. Holman {\it et al.}, Phys. Lett. {\bf 282B} (1992) 132;
S.M. Barr and D. Seckel, Phys. Rev. {\bf D46}
(1992) 539.}\foot{Usually the gauge symmetries of the
renormalizable model allow operators of dimension $5$ or $6$, but
discrete gauge symmetries can be invoked to push $d$ to larger values.  
In the M theory region of moduli space, $M_G \sim 3 \times 10^{16}$ GeV.}

For example, in an attempt to build a QCD axion model based on accidental
symmetries we must require that the gravitationally induced mass be
smaller than that coming from QCD.  In equations, we must have $\Lambda_{QCD}^2
M_G^{{\ha (d-4)}}>{f^{\ha (d-4) + 3}}$.  For an axion decay constant of
order $10^{11}$ GeV, this requires $d > 16$.  Similar restrictions apply
to majoron models.

\newsec{Cosmology}

Here we will make only the briefest remarks about cosmology in the M
theory region of moduli space.  The first thing to note is that the
large vacuum energy densities typical of many inflation models are
uncomfortably close to the eleven dimensional Planck scale.  This raises
the disturbing (or perhaps exciting) possibility that the inflationary
era can only be studied with the unknown machinery of quantum M theory.

Indeed, in the scenario we have presented in this paper for
nonperturbative physics and SUSY breaking, the natural scales of energy
density in the low energy four dimensional theory are all much lower
than $\Meleven$.  The vacuum energy density is of course moduli 
dependent, so
we can always imagine that inflation takes place in a region of moduli
space where the energy density is close to $\Meleven^4$.  We will then
have to deal with the \lq\lq cosmic overshoot\rq\rq problem described
by Brustein and Steinhardt\ref\brustein{R. Brustein
and P. Steinhardt, Phys. Lett.
{\bf B302} (1993) 196.}.  The initial energy density
of the system is much larger than the barriers that separate the
inflationary region of moduli space from the extreme weak coupling
region where string theory contradicts observation.
In \ref\bbs{T. Banks, M. Berkooz,
and P. Steinhardt,
Phys. Rev. {\bf D52} (1995) 705.} it was suggested
that this problem might be less severe than it had first appeared.
In a region of steeply falling potential, the moduli lose energy
exponentially in the distance covered by the trajectory on moduli space.
It requires a detailed knowledge of the lagrangian on moduli space to
determine whether the system really crosses the barrier into the weak
coupling region.

We also note that the natural candidates for inflatons in the M theory regime
are the bulk moduli.  They have self couplings which scale with powers
of $m_4$ so that the natural size of the forces restoring these moduli
to their equilibrium values is of the same order as gravitational
friction. 

Assuming that we can construct a satisfactory inflationary model, we
will certainly have to face a cosmological moduli problem.  
Many of the bulk
moduli have masses of the same order of magnitude as squarks in
strongly coupled heterotic string theory.  Despite the replacement of
$M_4$ by $\Meleven$ as the fundamental gravitational scale, $M_4$ (or
perhaps $m_4$) is the parameter which determines the couplings of the
moduli to ordinary matter.  We will have to borrow one of the existing
mechanisms for solving this problem\ref\cosmoduli{
L. Randall and S. Thomas, ``Solving the Cosmological
Moduli Problem with Weak Scale Inflation,"
Nucl. Phys. {\bf B449} (1995)
229, hep-ph/9407248; 
M. Dine, L. Randall and S. Thomas, ``Supersymmetry Breaking
in the Early Universe,"
Phys. Rev. Lett. {\bf 75} (1995) 398, hep-th 9507453;
D.H.Lyth, E.D.Stewart, {\it Phys. Rev.} {\bf D53}, (1996),1784, {\it
Phys. Rev. Lett.} {\bf 75}, (1995),201, hep-ph/9510204/9502417.}\bbs
or come up with a
new one.  Note that we also have a QCD axion with Planck scale decay
constant.  Most mechanisms (with the notable exception of \ref\linde{A.
Linde, ``Relaxing the Cosmological
Moduli Problem," Phys. Rev. {\bf D53} (1996) 4129.})
for solving the cosmological moduli problem will not help with the
axion.  However, the very existence of the moduli will change the nature
of the axion problem.  The very early universe will be cold and matter
dominated, so the usual analysis of axion history above the QCD phase
transition may not be relevant.

It should be clear furthermore that the cosmology of strongly coupled
heterotic string theory is considerably more complicated than models
that have been considered in the literature.  In addition to more or
less conventional bulk moduli and QCD axion fields, the model also has
boundary moduli.  These have mass of order the gravitino mass, but
couplings to matter suppressed only by powers of $\Meleven^{-1}$.  Their
reheat temperature is about $1$ MeV.  We also have scalar partners for
the axions, which will be a form of late decaying dark matter, and
probably have to have very small density at nucleosynthesis if they are
not to ruin classical cosmology.
The distributions of energy among the various scalar fields may lead to 
a rich and complicated cosmological scenario.  We will also have to sort
out the question of whether the QCD axion is dominantly a boundary
modulus in generic regions of moduli space in order to embark on a
detailed study of the cosmology of M theory.

\newsec{Conclusion}

Strongly coupled heterotic string theory retains most of the attractive
features of the weakly coupled region but provides a better fit to the
parameters of the real world.  There is no longer a discrepancy between
string theory and supersymmetric coupling unification.  In the strongly
coupled region there is always a QCD axion and the strong CP problem is
resolved.  The axion decay constant violates cosmological bounds, but we
view this as a challenge rather than a definitive failure of the theory.
Indeed, the most serious phenomenological problem of string theory, in
any region of moduli space is the cosmological moduli problem.
Several solutions to this have been proposed, but
Linde's seems to be the only one which could resolve the axion problem.
The axion is of course also an attractive dark matter candidate.
If $h_{1,1} > 1$, the theory predicts a number of essentially stable
axionlike particles with Compton wavelengths of astrophysical magnitude.
If boundary moduli contribute to the QCD axion then we
will certainly have at least one of these particles.
Observations measuring the number of such light axions would be of the
utmost interest.  They would amount to measurements of topology of the
six compactified dimension.  Alternatively, if no such particles are
found, and if there are boundary contributions to the QCD axion, the
entire M theory region of moduli space would be ruled out.

We have also proposed a scenario for SUSY breaking in the strong
coupling region.  The fundamental reason for the 
discrepancy between the Planck scale and the unification scale is the
existence of a
fifth \lq\lq large \rq\rq dimension an order of magnitude larger than
the unification scale.  As a consequence, certain fields of the theory
exhibit an approximate $5$ dimensional supersymmetry which is broken by
terms of order inverse powers of the radius of the fifth dimension.
There are $h_{1,1}$ chiral superfields in the low energy four
dimensional theory which descend from vector multiplets in five
dimensions.  The axions are the imaginary parts of these fields.
Approximate $N=2$ SUSY produces an approximate {\it no scale} scenario
for SUSY breaking, in which the F terms of the axion multiplets are the
order parameters.  The R symmetry breaking which triggers SUSY breaking
comes from nonperturbative physics on the strongly coupled boundary.
We argue that in this scenario squark degeneracy naturally arises to
leading order in the inverse radius of the fifth dimension.
The scenario also leads to $h_{1,1}$ scalar axion superpartners, with
masses of order $100$ GeV and Planck scale couplings to matter.  These are
a form of late decaying dark matter and are constrained
by classical cosmology.

The scenario is unacceptable as it stands.  If we make the natural
assumption that strongly coupled physics does not introduce any small
parameters into the superpotential, then we predict the SUSY breaking
scale to be much too large.  Otherwise, we must resort to the sort of
Kahler stabilization of some of the moduli that we advocated in
\banksdine\ for the regime of weakly coupled string theory.  
Apart from this, we must also invoke higher order terms in the expansion
in the inverse radius of the fifth dimension to explain the
stabilization of the radius.  This is precisely the sort of procedure
that was criticized in \dineseiberg.  Here however the expansion
parameter is only of order $0.1$.  It is plausible then that the
expansion breaks down for the values of the moduli at which the minimum
is achieved.  It is also reasonable to use the expansion as evidence for
the existence of a SUSY breaking minimum (though not of course to
understand why the cosmological constant is zero).  However, we do not
see how to save the prediction of squark mass universality which follows
{}from the no scale structure at large $\Releven$. 

It is fairly clear from this discussion that we do not
yet understand the mechanism of SUSY breaking in the M theory regime.  
We suspect that this may
be closely connected with another phenomenological issue that has not
yet been explored, the quark mass matrix.  Most successful theories of
the quark mass matrix are based on horizontal symmetries.  In string
theory, an attractive origin for horizontal symmetries has been
suggested by a number of authors\ref\fayet{
J.L. Lopez
and D.V. Nanopoulos, Nucl. Phys. {\bf B338} (1990) 421;
L. Ibanez,
Phys. Lett. {\bf B303} (1993) 55; L. Ibanez and G.G. Ross,
Phys. Lett. {\bf B332} (1994) 100;
A.E. Farragi, Phys. Lett. {\bf B274} (1992) 47,
Phys. Rev. {\bf D47} (1993) 5021;
A.E. Faraggi and E. Halyo, Phys. Lett. {\bf B307} (1993) 305;
Nucl. Phys. {\bf B416} (1994) 63; P. Bineutry
and P. Ramond, Phys .Lett. {\bf B350} (1995) 49.}.  
They originate as $U(1)$ gauge
symmetries which have Fayet-Iliopoulos D-terms.  We feel certain that
the dynamics of cancellation of the D-term will influence the breaking
of supersymmetry and the stabilization of the moduli.  Perhaps it will
help to resolve some of the puzzles we have uncovered.

In the long term, if the M theory region of moduli space has anything to
do with the real world, the most striking feature of its phenomenology
will be the low scale at which interesting gravitational phenomena
become accessible.  At energies of order $10^{15}$ GeV,
``experiments" will reveal an extra bosonic dimension of spacetime, and
discover that some of the degrees of freedom live on \lq\lq the other
wall of the world\rq\rq .  At energies one or two orders of magnitude
higher we will encounter true quantum mechanical manifestations of
gravity and find out what M is.  We have already indicated that the low
scale of gravitational phenomena forces us to envisage discrete
symmetries which forbid the leading gravitational corrections to the 
standard model.  It
is to be hoped that further study will reveal interesting signatures of
M theory that can be probed at low energies, or in the early universe.

\centerline{{\bf Acknowledgements}}
\noindent
We thank M.Berkooz, R.Leigh, Y.Nir, A. Rajaraman,
N. Seiberg, S. Shenker, Y. Shirman, L. Susskind, S. Thomas, P. Townsend
and E. Witten for conversations.  The work of M.D. was supported
in part by the U.S. Department of Energy.
The work of T. Banks was supported in
part by the Department of Energy under
grant $\# DE-FG0296ER40559$.

\listrefs
\bye

\end